# The Reverend Walter Bidlake of Crewe

Jeremy Shears

## Abstract

The Reverend Walter Bidlake, MA, FRAS, JP (1865-1938) was vicar of Crewe for some 21 years. A member of the BAA, he was a keen amateur astronomer with an interest in celestial photography. At the request of another well-known BAA member, T.H.E.C. Espin, Bidlake took photographs of the night sky in support of Espin's search for dark nebulae in the Milky Way. This paper describes Bidlake's astronomical activities and life, including a high profile libel case he brought.

## Introduction

As a regular commuter from Crewe on the West Coast line, I was not aware of any particular astronomical connection with the railway town. Therefore I was surprised to see a reference to the Reverend Walter Bidlake of Crewe (Figure 1) in a 1922 paper (1) written by the eminent BAA member, the Reverend T.H.E.C Espin (1858-1934), and entitled *Dark Structures in the Milky Way*. On reading further, I learnt that Bidlake had taken photographs of the night sky from Crewe which Espin had used in this research. Central Crewe is no longer the obvious place for conducting astrophotography, especially considering the invasive sky glow from the station and its associated marshalling yards which are permanently floodlit (2). So what was Espin's project and who was the Reverend W. Bidlake?

## Espin's *Dark Structures in the Milky Way*

On 16 January 1898 Espin was sweeping for red stars in Perseus when he "passed suddenly from the starry background into what appeared to be a cloud. Although the night was very clear, yet I felt convinced it was a cloud, and continued with my sweep. At the end of the sweep the telescope, as usual, was moved 40' south and the return sweep made. I again came on the unusual obscuration" (3). Similar observations were made later that night and again on 24 and 25 January. Espin notified J.K.E. Halm (1866-1944) of the Royal Observatory Edinburgh of his findings and Halm picked up the object on 17 February. Several years later, in 1911, having completely forgotten about his earlier observations, Espin stumbled across the spot again. This time he had to hand photographs of the region taken by Max Wolf (1863-1932) of Heidelberg University. Espin tracked down the position of his object on the photographs and found that they "show it as a hole in the Milky Way". This stimulated Espin's interest in apparently starless regions of the Milky Way. The nature of the so-called "dark nebulae" was receiving much attention from astronomers around the world at the time due to the appearance of wide-field photographs of the Milky Way published by Wolf, W.S. Franks (1851-1935), E.E. Barnard (1857-1923) and others. Espin described some of these objects in a paper in 1912 (4), which as had been suggested by others, he interpreted as being due to absorption of light from distant stars by foreground matter.

It is of course Barnard's name that is nowadays chiefly associated with the dark nebulae, especially his fine photographs of the Milky Way showing these features taken with the Bruce astrograph on Mount Wilson. In 1919 he published his catalogue of 182 such objects, later extended to 346 objects (5), which is still used today. Espin's object in Perseus is listed in the catalogue as B 20 (RA 04 h 37.1 min, Dec. +50° 58', J2000.0) and Barnard cited Espin's 1898 observation of it:

"In S part of larger, relatively vacant area nearly 1° in diam. This is the dark object mentioned by Espin in Monthly Notices, 58, 334, 1898. It is close north to a small group of faint stars and is 6' or 8' in diameter. In a somewhat larger vacant space. It is not so definite as B 15."

Modern images of the region are shown in Figure 2.

It might have been the appearance of Barnard's catalogue that reignited Espin's interest in dark nebulae, for later in the same year, 1919, recognising the usefulness of wide-field photography in identifying these objects, Espin contacted the Director of the BAA Photographic Section, F.W. Longbottom (1850-1933) (6), to ask if Section members could obtain photographs of the sky between Dec. +30 and +60° to assist him in his research. The project was duly launched and Longbottom noted progress in his annual reports between 1919 and 1922 (7).

The timing of the project was opportune since interest in celestial photography amongst BAA members, several of whom had been away on active service, was making a resurgence following end of the First World War and during which it had become increasingly difficult to obtain the required photographic materials. It was also helped by high quality photographic lenses coming onto the military surplus market at affordable prices. Moreover, the hostilities had resulted in the development of more sensitive photographic emulsions which were better suited to astronomical applications.

In 1922 Espin published (1) his own list of 202 dark nebulae based on an examination of photographs by Barnard as well as those submitted by "Mr. H. Ellis of Lyme Regis, Rev. W. Bidlake of Crewe, Mr. J.M Offord of West Ealing, Mr. Lawrence Richardson of Newcastle-upon-Tyne, Mr. F.W. Longbottom of Chester, Mr. Evens and Mr. Milburn". This had the same title as his paper of 10 years earlier: *Dark Structures in the Milky Way*. The paper contained no illustrations and the current whereabouts of the photographs taken by Bidlake and the others is unknown; it is entirely possible that they have been lost.

**Bidlake's early life and ecclesiastical career**

Walter Bidlake was born in Wellington, Shropshire, on 15 August 1865. His parents were John Bidlake (1833-1890), a solicitor, and Frances Corser Barker (1842-1930); they had a total of 4 children (8).

Walter was educated at the Leys School in Cambridge and went up to Exeter College Oxford, graduating BA in 1887 and MA in 1891 (9). He trained for the ministry at Leeds Clergy School and was ordained Deacon in 1888 and Priest in 1889. He spent his career in the Diocese of Chester, commencing with Curacies at St. Oswald, Chester (1888-90; see also Figure 3) and St. Paul, Coppenhall, Crewe (1890-96). He was then, in succession, vicar of Ellesmere Port (1896-1902), Christ Church Crewe (1902-23; Figure 4) and Neston, on the Wirral (1923-1938) (10). During the Neston period he also became a Canon of Chester Cathedral.

Bidlake married Lucy Mirion Atkinson (1875-1908), the eldest of twelve children of James Atkinson JP in 1899 (11). Life in Crewe revolved around the London & North Western Railway Company (LNWR). James Atkinson was employed by the company as a surgeon and he went on to become town's first Mayor in 1877. The Atkinson family were well-known in Crewe social circles and a part of Christ Church, which itself had been built by the railway company (12), became known informally as "the Atkinson chapel" as a result of several stained glass windows having been dedicated to various members of the family. Of course, a mere three years after the marriage Bidlake was to become vicar of Christ Church.

Walter and Lucy had a daughter, Lucy Mary, in 1905 (d 1962). Three years later twins were on the way, but tragedy struck: both they and their mother died in childbirth, leaving Walter to bring up his daughter on her own. In 1931 Lucy Mary (familiarly known as Molly) married Gerald Victor Haldane Unwin (1897-1976) at Neston with the wedding ceremony being conducted by Walter Bidlake as he was the vicar of Neston at the time.

Apart from his clerical duties, Bidlake was a Justice of the Peace and an elected Councillor for Cheshire County Council, representing Crewe around the time of the First World War (13). He was a keen cricketer (14) and a member of the Oddfellows Lodge (15). He also actively supported the 1st Crewe (Christ Church) Troop of Boy Scouts which was based at Christ Church when he was vicar. The Troop (Figure 5) was started in February 1908, within a few weeks of a visit to Crewe by Lord Baden-Powell (16).

## Astronomical interests

Bidlake's first published astronomical observation was of a "brilliant meteor" he saw from Crewe on the evening of 24 September 1894 which he wrote about in a letter to the *English Mechanic* (17). Later he owned a 3½ inch (9 cm) refractor and a 5 inch (12.5 cm) refractor by Broadhurst, each equatorially mounted (18). He was elected as a member of the BAA on 26 February 1913 (19) and as Fellow of the RAS in 1914 (20), having been proposed for the latter by W.E. Plummer (1849-1928) who was Director of the Liverpool Observatory at Bidston Hill, near Birkenhead (21). Both Bidlake and Plummer were active in two local astronomical societies, the Liverpool Astronomical Society and the Astronomy Section of the Chester Society of Natural Science & Art (22), and were evidently well acquainted. Fred Longbottom, who lived in Chester, was also a leading light in both Societies and was a friend of Bidlake's. It was Longbottom who proposed Bidlake for BAA membership (23) and it might have been Longbottom's infectious enthusiasm for celestial photography that encouraged Bidlake to join the BAA Photographic Section and to take up the branch of astronomy in which he became most active.

Apart from references to Bidlake's work on celestial photography under the auspices of the BAA Photographic Section and some observations submitted to the Meteor Section, there is little record of other work that he might have conducted. The only other detailed astronomical observation he published was of the annular solar eclipse of 8 April 1921, which he observed from St. Cyr, near Paris. He was actually located about 2 miles south of the central line. Observing with binoculars, "At maximum the Moon was surrounded by a very thin circle of sunlight broken by a number of black dots, which disappeared almost as soon as they came in view" (24).

By contrast to Bidlake's published astronomical work, rather more column inches relate to two legal cases he brought and it is to these that we will now turn.

**Cases of deception, libel and blackmail**

Not long after taking up his position at Christ Church Crewe, Bidlake became the victim of a confidence trick, the essential details of which were carried in the newspapers. The story was that a lady had come to the vicarage in early 1903 seeking alms and requesting the vicar to visit her dying child at her home. He sent her away with some money and later set off from the vicarage to visit them, only to discover that the name and address that she had given him were false. Consequently, he reported the matter to the Police and insisted that the case be brought forward against her "because he thought that the clergy were too often duped" (25). The lady, whose real name was Rhoda Hughes, was arrested and duly appeared in front of the magistrate at Crewe, where she was bound over.

Bidlake's second legal case also illustrates the vulnerability of the clergy resulting from the public nature of their vocation. However, this case was much more serious, involving libel and possibly blackmail heralded by the salacious headline in *The Post* on 28 Nov 1915 (26):

## BLACKMAIL ALLEGED BY VICAR.

————

## LADY WHO CLAIMED HE WAS TO MARRY HER.

The story was also carried by other papers (27), with the most detailed article appearing on the front page of *The Manchester Courier* on 29 November 1915. This set out the nature of the charges which were that the defendant, Emma Mary Fulham (aged 52) of Crewe (28), "unlawfully wrote and published a false and defamatory libel" against Bidlake. Fulham, a widow, had applied to Bidlake in 1913 for parish work, which she was duly given, but shortly afterwards left for Hereford. Sometime later Bidlake, by then a widower of some 8 years, received a note from Hereford in which the correspondent stated that Fulham had given his name as a reference. Shortly afterwards he received a bill for her board and lodgings at Hereford, with a note explaining that it had been sent to him as her "affianced husband". Sometime later Fulham returned to Crewe to find Bidlake. She began to write letters to him, as well as to other clergy in the Crewe area and the Bishop of Chester, about their betrothal. She even requested the local clergy to prepare their churches for their forthcoming marriage.

Things were clearly escalating and Bidlake felt he needed to act. A summons was served on Fulham who promptly went round to Bidlake's vicarage and "threw at him an envelope" which contained the following letter, no longer merely citing an engagement, but now a marriage (29):

*To my husband, Walter Bidlake,*

*You wretch, don't think I am to be bluffed by the enclosed. I shall not appear at any court. I am your lawful wife, Mary Bidlake.*

*Emma Fulham died a natural death on February 13th, 1913, when you married me yourself at Nantwich Road. You cannot mock God.*

*Wooldridge indeed* [meaning the magistrate who signed the summons]. *Tell him he is talking through his hat. You may keep the precious document for another souvenir. I could kill you. You, a priest, to dare to play so. Do you think I am an idiot? I will see you and Woodridge hanged before I go. Imagine a man of his position talking such a piece of piffle to me.*

*I wrote to the Bishop this morning. You will have to acknowledge me now. Hughes* [a local curate] *is a champion actor. You ought to be on the boards. Parsons, indeed. By heavens ! you take the cake.*

*Your fighting wife,*

*MARY BIDLAKE*

The Crewe magistrate who issued the summons also received "hundreds of letters and postcards from the accused, and some he refused to take in". Eventually, Fulham was arrested in Birmingham, when she was quoted as saying: (29)

"I am not Mrs. Fulham, I am Mary Bidlake, and we were joined under a covenant before God in Nantwich Road on February 13th, 1913, which was more binding than if we had been married before God's holy altar, or at any registry office. It was ridiculous to bring such charges against me. Why does not Mr. Bidlake come here and make it himself. He is my husband, and should be here. The Bishops know all about it".

As a result of the preliminary hearing at Crewe, Fulham was committed to appear before the Chester assizes the following February (1916). The trial, in which she was charged with criminal libel against Bidlake, was also reported widely in the press with headlines such as (Figure 6): (30)

### VICAR LIBELLED.

---

### Remarkable Crewe Case.

### Insane Lady Church Worker.

However, the case collapsed on the first day. A true bill was returned by the Grand Jury, indicating that Fulham had a case to answer (31), but there immediately followed a submission by Dr. Priest, the prison doctor at Strangeways Prison in Manchester, where she had been held awaiting trial. The doctor had had Fulham under observation since her arrival at the prison in November and had come to the conclusion that "she was insane and unfit to plead" (32). The jury duly found her insane and the Judge, Mr. Justice Lush (33) (1853-1930, Figure 7), concluded by saying that there were no grounds for any of her claims against Bidlake and ordered her detention during His Majesty's Pleasure. Thomas Artemus Jones (1871-1943, Figure 8), the Barrister acting for Bidlake (34), applied that Bidlake should not have to bear the costs of the prosecution, noting: (32)

"The reverend gentleman was a well-known vicar and public worker in the county. He was reluctant to take proceedings, but was obliged to do so to protect himself."

The Judge responded by agreeing that Bidlake "could not take any other course. He was quite right. The cost of the prosecution will not have to be borne by the vicar".

And thus the case was over, presumably to Bidlake's great relief. Being innocent of the claims made against him, and in spite of the sensational reporting and great public interest, a scandal was averted and his career was apparently unaffected.

## Later life

After some 21 years at Christ Church, Crewe, Bidlake was appointed to the living of Neston, Cheshire, where he became vicar of the Church of St. Mary & St. Helen on 23 October 1923 (35). He remained in that position until early 1938 (36), moving to Nantwich where he passed away on 21 November (37). There is no further mention of his astronomical exploits until his passing is noted in the BAA Annual Report for the 1938/39 session, at which point he was still a member (38).

Although the membership of the BAA was drawn from many social backgrounds, members of certain professions were particularly well-represented in its early days, such as medical doctors, military officers and the clergy. Possessing sufficient financial security and disposable leisure time, they were able to develop their own hobbies and interests. Bidlake was perhaps typical of such members, equipping himself with the necessary telescopes and cameras to pursue his hobby. Whilst he might not have pursued astronomy to the same degree as his fellow clergyman, T.H.E.C. Espin, he nevertheless had a keen interest in the subject and contributed to the observational and practical activities of the BAA.

## Acknowledgements

I am most grateful to several people who have assisted me in this research. Denis Buczynski responded to a request I placed on the BAA web site for an image of the Barnard 20 region and allowed me to reproduce it in this paper. Gerald Newbrook provided details about the development of Crewe in the nineteenth and twentieth centuries. He brought to my attention the photograph of the 1st Crewe (Christ Church) Scouts, which in turn he had been given by David Walters. I thank both gentlemen and the South West Cheshire District Scout Association for granting permission to reproduce the photograph. George Bidlake, great grandson of Walter Bidlake's brother George, and Sarah Unwin, wife of Bidlake's grandson Hugh, have provided much information about the Bidlake family history. Tony Kinder provided information about Bidlake's BAA membership and probate. Janet Rossiter helped with information about Bidlake's time at Neston.

This research made use of scanned back numbers of the Journal, which exist largely thanks to the herculean scanning efforts of Sheridan Williams, and the English Mechanic, supplied by Eric Hutton. The NASA/Smithsonian Astrophysics Data System, the British Newspaper Archive (British Library), the Cheshire County Archives, the online archives of the Family History Society of Cheshire (Crewe Group) and the Digitized Sky Survey, produced at the Space Telescope Science Institute under U.S. Government grant NAG W-2166, were also used in the preparation of this paper.

I am grateful to the referees, Mike Frost and Stewart Moore, for their helpful comments.

extended in 1864 by addition of aisles, by the tower in 1877, by the chancel in 1898 and by the chapel in 1906. During Bidlake's ministry the old peal of eight bells was recast and expanded to a peal of ten. In the early 1970s, wood rot was discovered in the nave roof and the church was partially demolished in 1977. Information from Family History Society of Cheshire, Crewe Group: http://www.scfhs.org.uk/scfhs/christchurch_crewe_intro.html.

13. Bidlake is listed as being elected to Cheshire County Council in 1913 and 1919 - the record is incomplete, so he might have been elected at other times too.

14. A one time, whilst at Ellesmere Port, Bidlake was Vice-Captain of the local Whitby & District Cricket Club; Cheshire Observer 29 April 1899, pg 8.

15. He was elected to the Marquis of Westminster's Lodge (Manchester Unity) of Oddfellows in 1896; Cheshire Observer, 21 November 1896. Later he was Chaplain to the Cheshire Grand Lodge of Mark Master Masons of England and Wales.

16. General Baden-Powell visited Crewe on 23 January 1908 to explain his idea of Scouting at a public meeting. Crewe was one of the first towns in the Country to form a Scout Troop. South West Cheshire District Scout Association. http://www.southwestcheshirescouts.org.uk/index.php/district-history.

17. "This evening about seven o'clock I observed a brilliant meteor travelling across the sky in an apparently straight line from east to almost due west....". English Mechanic, 1540, 134 (1894).

18. The 5-inch refractor had a focal length of 60-inches (152 cm). It was put up for sale for £30 in 1921. English Mechanic, 2933 (1921).

19. JBAA, 23, 251 (1913).

20. Elected on 8 April 1914. MNRAS, 74, 34 (1914).

21. An obituary of Plummer and details of the Bidston Observatory can be read in Hollis H.P., MNRAS, 89, 320-323 (1929).

22. According to the records of the Chester Society of Natural Science & Art, Bidlake made at least one presentation at an Astronomy Section meeting, on 15 December 1915, his topic being "the Stars in Winter".

23. JBAA, 23, 219 (1913).

24. Bidlake W., English Mechanic, 2457, 299 (1921).

25. The legal proceedings are mentioned in The Manchester Evening News, 26 March 1903, pg 2, under the title "Imposing on Clergy".

26. The story was carried on page 2.

27. Other papers included the Liverpool Echo of 25 November 1915: "A Crewe vicar complains of libel by a widow".

28. Fulham's address was given as Derrington Street, Crewe, which no longer exists. Local historian, Gerald Newbrook, points out that on the 1911 OS map there is an L-shaped road and one section is named Derrington Street and the other is Derrington Avenue. Both are now named Derrington Avenue.

29. Quoted in a report in The Manchester Courier, 29 November 1915.

30. Manchester Evening News, 23 February 1916.

31. The Grand Jury would evaluate charges and return what was called a "true bill" if the charges were to proceed. England abandoned Grand Juries in 1933 and instead uses a committal procedure.

32. Liverpool Echo, 25 February 1916.

33. Sir Charles Montague Lush. Educated at Westminster School and Trinity Hall Cambridge. He was called to the Bar in 1879 and took silk in 1902.

34. Later in 1916 Artemus Jones was on the team defending Sir Roger Casement (1864-1916), who was tried for treason in connection with Irish nationalist activities. Following an unsuccessful appeal, Casement was hanged at Pentonville Prison on 3 August 1916. Jones later went on to be a Judge. In 1910 Jones had brought his own libel case. The Sunday Chronicle had published a satirical sketch about one Artemus Jones, a fictional Peckham church warden, who had gone to France with a woman "who was not his wife", Jones (not from Peckham and not a church warden) complained and was awarded the sum of £1,750 in libel damages. He satisfied the House of Lords that reasonable people might conclude that the defamatory words referred to him.

35. Bidlake's new position was publicly announced in The Evening Telegraph of 30 July 1923. Whilst at Neston he lived at 'The Manse', Parkgate Road.

36. I am not certain exactly when he retired from his position at Neston. His successor, Rev. McCauley Bennett, became vicar on 11 February 1938, but there might have been interregnum.

37. Probate records show that Bidlake was living at Hillfield House, Wellington Road, Nantwich, Cheshire when he died. His effects of £3809 8s 10d passed to his daughter.

38. JBAA, 49, 374 (1939). Who Was Who, 1929-1940 edition, lists Bidlake's recreations as fishing and golf, with no mention of astronomy.

39. David Walters' grandfather, Harold H. Walley, is seated second from right.

40. Jones T. A., Without My Wig, publ. Brython Press, Liverpool, 1944.

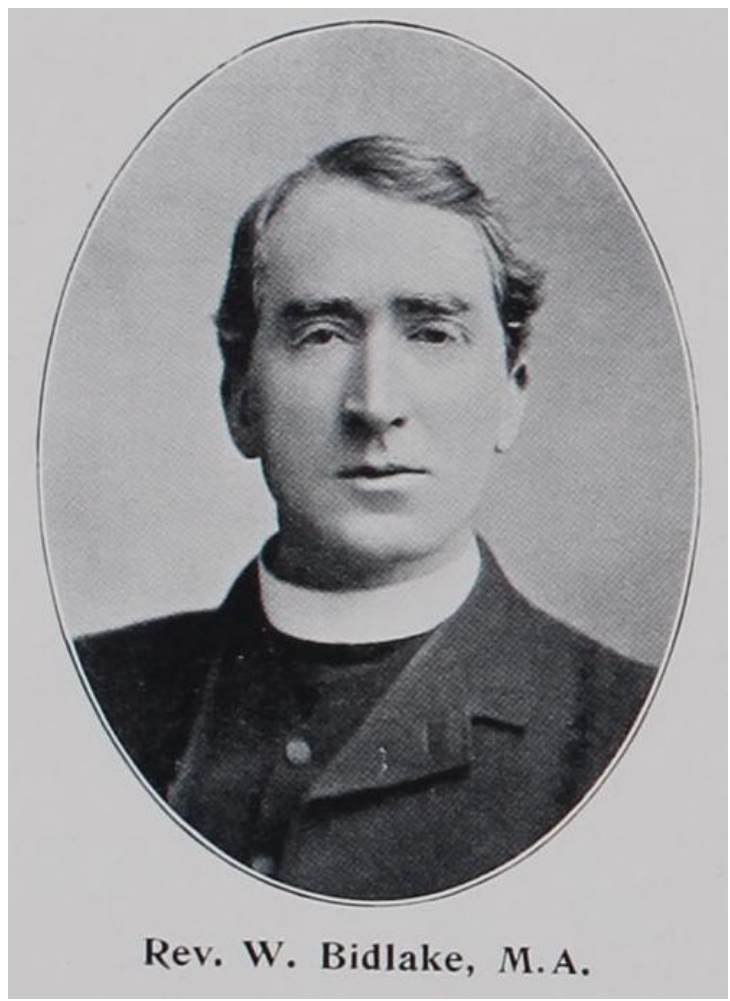

Rev. W. Bidlake, M.A.

Figure 1: The Reverend Walter Bidlake (1865-1938)

From reference (9)

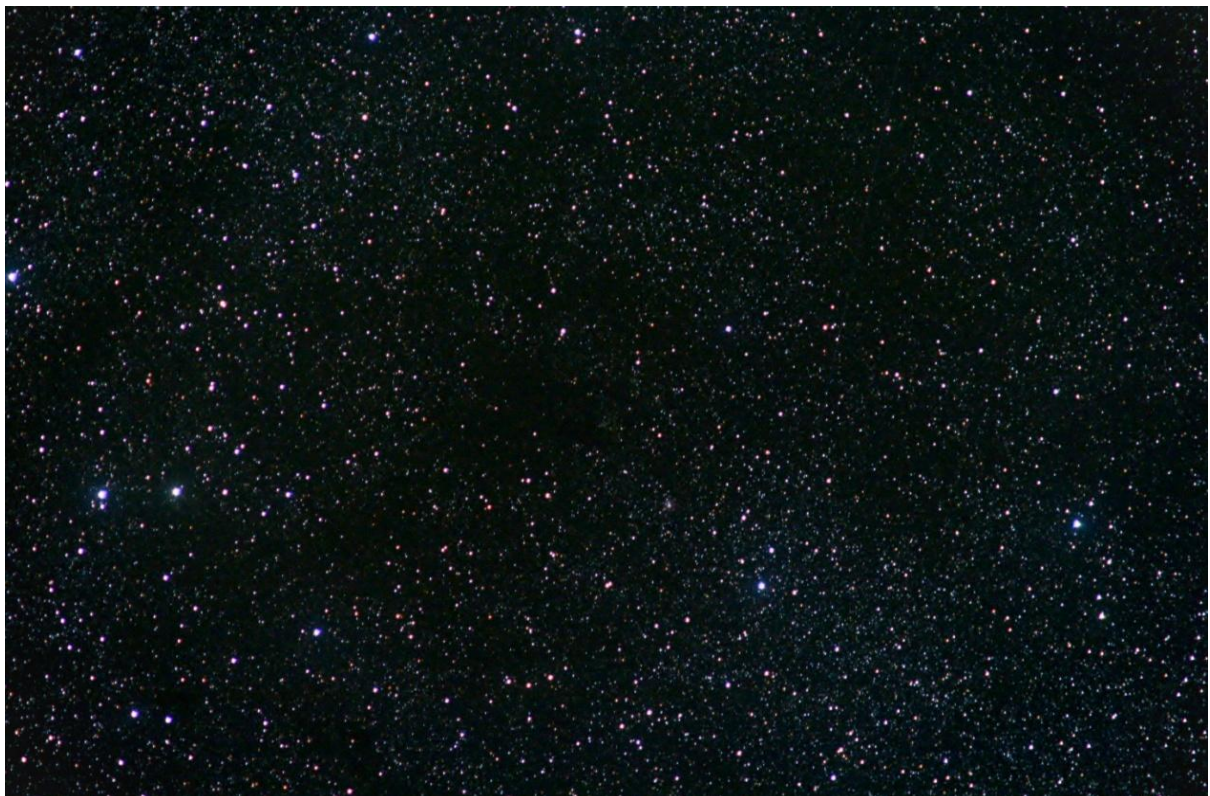

(a)

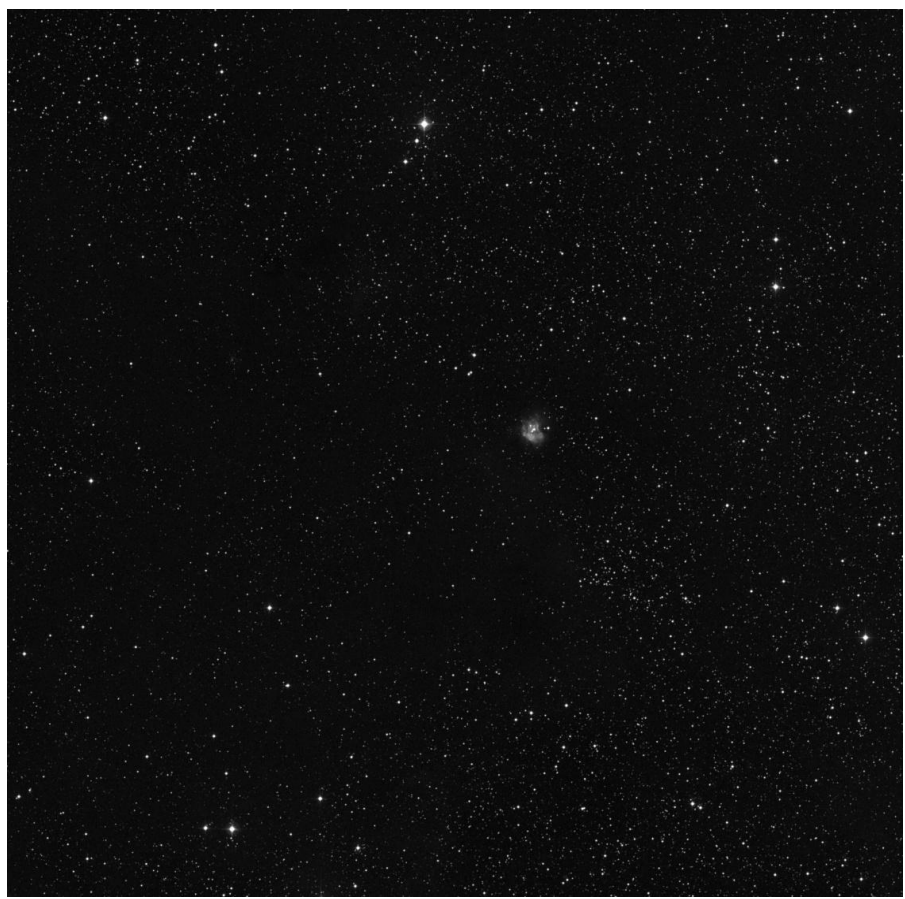

(b)

Figure 2: The region of Barnard 20 in Perseus

(a) DSLR image Denis Buczynski. B 20 covers the central part of the picture. North left, West up. Exposure 475 seconds on Canon 400D camera at ISO 1600 with a 200mm Mirage auto reflex lens at f/4 on 2013 Jan 01 19:06 UT

(b) Digitized Sky Survey image of B20, 60' x 60', orientation as before. B 20 extends over much of left side of the field (and beyond). The nebula to the right of centre is the HII region, Sh 2-211 (not visible in the DSLR image due to its insensitivity to Hα light). Below and to the right of Sh 2-211 is the open cluster Berkeley 67

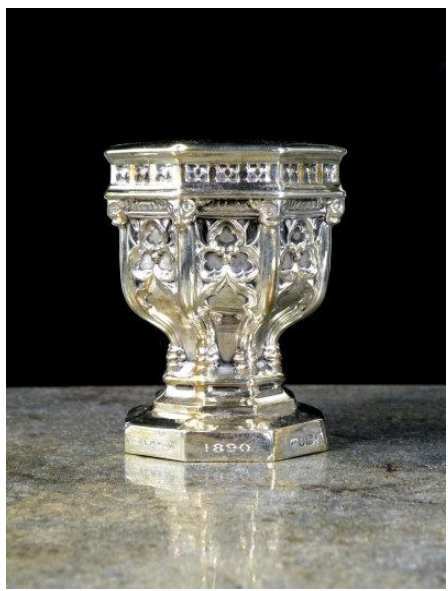

Figure 3: A silver presentation miniature engraved to the base 'Rev Walter Bidlake, A gift from S. Oswald's Chester, November 1890'

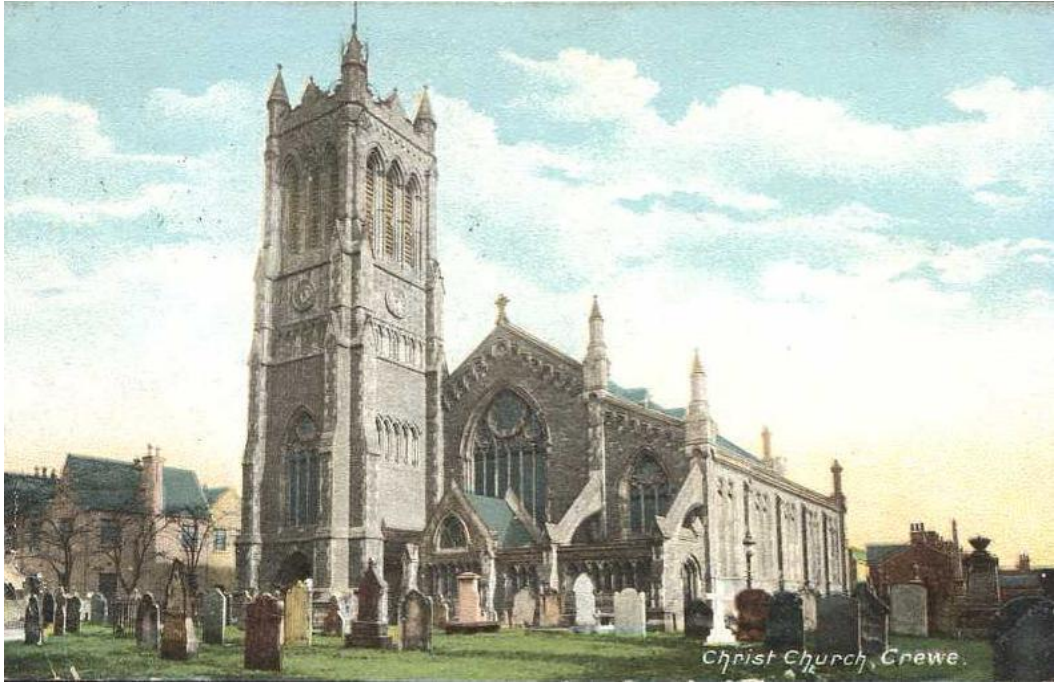

Figure 4: Christ Church Crewe in 1905

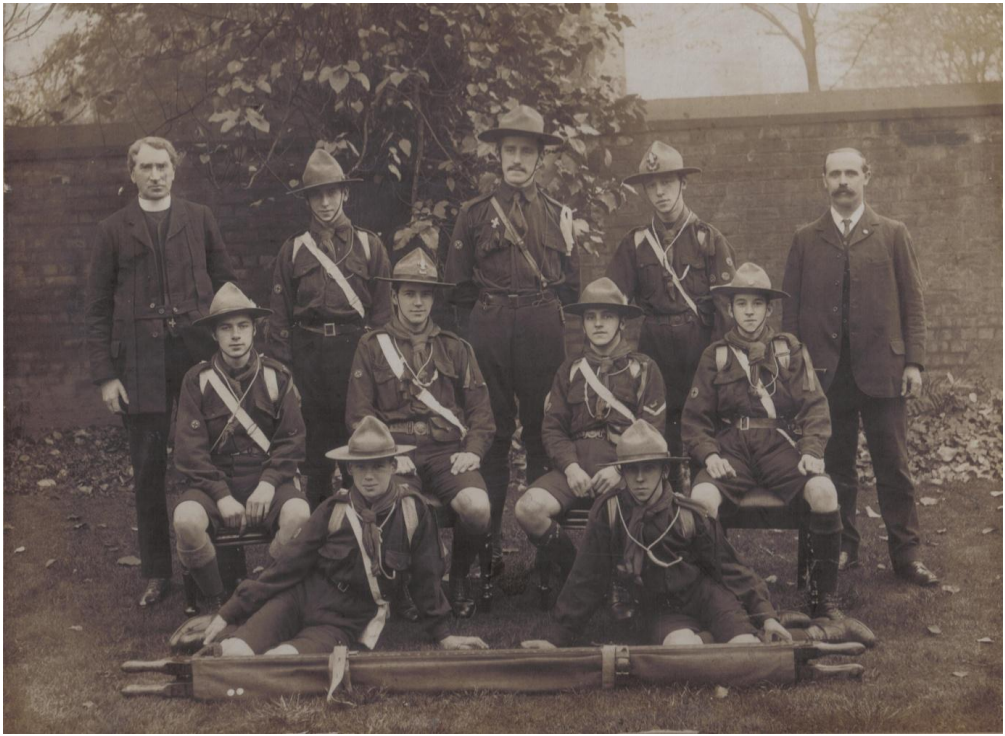

Figure 5: The Reverend Bidlake and the 1st Crewe (Christ Church) Scouts ca. 1910

The Scoutmaster, centre of back row, was a local Accountant named Wilmot Welch, who had formed the Troop in 1908. The photograph is believed to have been taken in the grounds of Christ Church Vicarage. The man on the right has not been identified. Image courtesy of Gerald Newbrook, David Walters (39) and the South West Cheshire District Scout Association archive

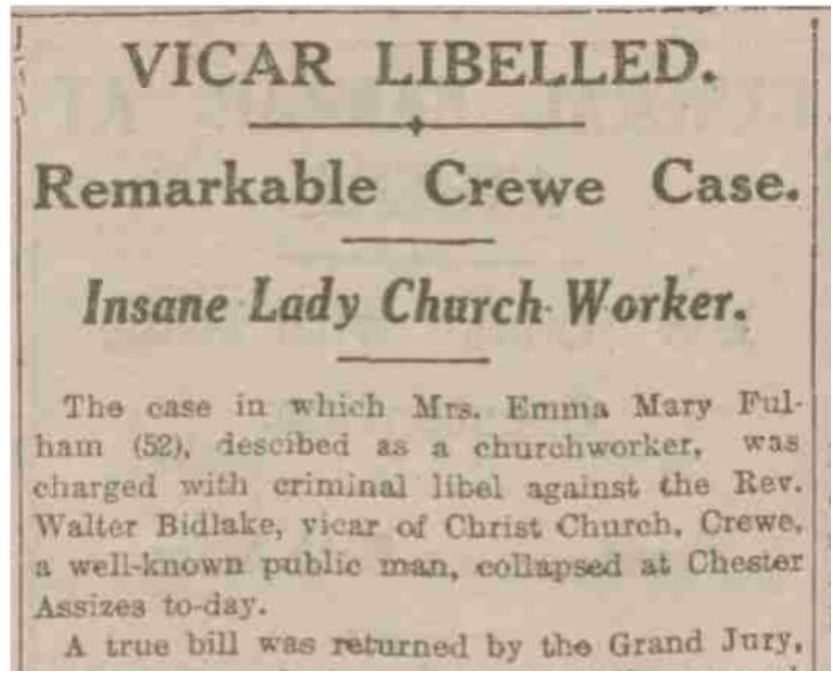

Figure 6: The case of Bidlake v. Fulham as reported in the Manchester Evening News on 16 February 1916

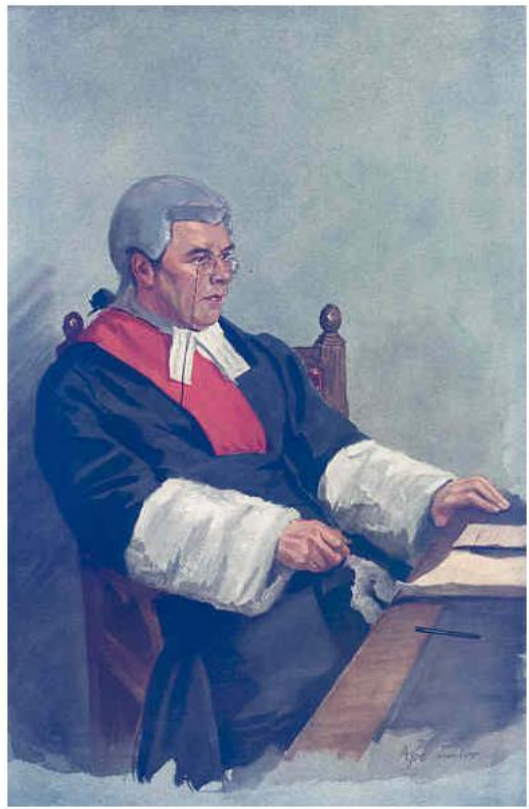

Figure 7: Mr. Justice Lush in 1911 (1853-1930)

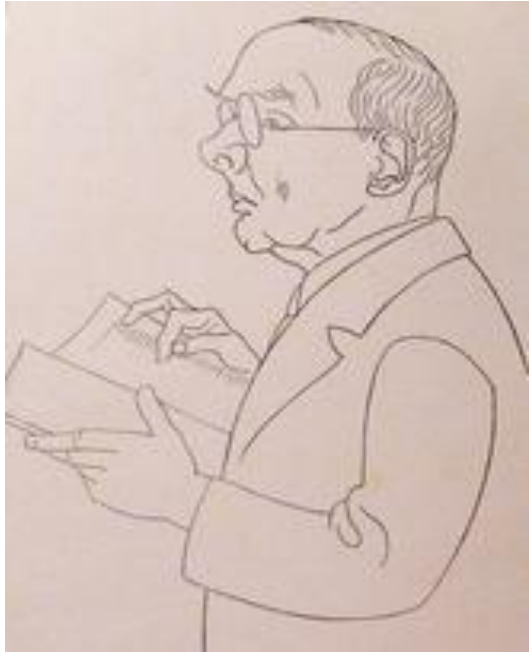

Figure 8: Thomas Artemus Jones (1871-1943) in later life

Caricature from his autobiography, *Without My Wig* (40)